\documentclass [12pt]{article}
\usepackage {graphicx}
\usepackage {longtable}

\begin{document}

\begin{center}
\LARGE\bf{Octonions and vacuum stability}
\end{center}

\bigskip

\begin{center}
\textbf{M.V. Gorbatenko}
\end{center}

\bigskip

\begin{center}
Russian Federal Nuclear Center - VNIIEF, Sarov, Nizhni Novgorod region 
\end{center}

\begin{center}
E-mail: .\underline {gorbatenko@vniief.ru}
\end{center}

\bigskip

\subsubsection*{Abstract}

\bigskip

The paper addresses one of nontrivial octonion related facts. According to 
paper gr-qc/0409095, the most stable space-time state is the one described 
by real Dirac matrices in 11-dimensional space of signature $1\left( { -}  
\right)\& 10\left( { +}  \right)$. The internal subspace is 7-dimensional, 
and its stability is due to a high ``zero'' energy packing density when 
using an oblique-angled basis from fundamental vectors of lattice $E_{8} $ 
for the spinor degrees of freedom. The nontrivial fact consists in the 
following: Dirac symbols with octonion matrix elements can be used to 
describe states of the space of internal degrees of freedom if and only if 
the space corresponds either to stable vacuum states or states close to the 
just mentioned ones. The coincidence of the internal space dimension and 
signature for absolutely different and independent approaches to the 
consideration of this issue seems to predetermine the internal space vacuum 
properties and the apparatus, which is able to constitute the basis of the 
unified interaction theory.

\bigskip

\section*{1. Introduction}

\bigskip

4-dimensional Riemannian space with metric tensor $g_{\alpha \beta}  $ is 
considered. The Greek letters take on values $\alpha ,\beta ,... = 0,1,2,3$. 
The metric tensor is assumed nonsingular, so Dirac symbols (DS) $\gamma 
_{\alpha}  $ according to

\begin{equation}
\label{eq1}
\gamma _{\alpha}  \gamma _{\beta}  + \gamma _{\beta}  \gamma _{\alpha}  = 
2g_{\alpha \beta}  
\end{equation}

\noindent
can be introduced at every point. This paper considers algebraic properties 
of DS at some spatial point. The coordinate system is assumed to be locally 
Cartesian and tensor $g_{\alpha \beta}  $ equal to

\begin{equation}
\label{eq2}
g_{\alpha \beta}  = diag\left( { - 1,1,1,1} \right).
\end{equation}

In the DS theory there are a number of problems, the solution to which 
directly affects the physical interpretation of DS involving constructions, 
however, such that there is no complete understanding in regard to their 
solution method. Mention two of them.

It is well known that symbols $\left\{ {\gamma _{\alpha} }  \right\}$ can be 
realized as Dirac matrices (DM) above any number field (real, complex, 
quaternion numbers) as well as above the octonion body. The realization in 
the form of square matrices $4 \times 4$ is meant. On the other hand, DS can 
be realized in the form of real square matrices $N \times N$, $N \ge 4$. Of 
interest is the question: What is the relation between these two realization 
types? In particular, what are the characteristics of the subspaces of 
internal degrees of freedom that are introduced additionally in each 
complication of the number body used? 

For physics, the octonion realization is of a special interest, in 
particular, for the reason that using any number body except for the 
octonion one does not allow us even to pose the question of explanation of 
the irreversibility of actual processes on the basis of time-reversible 
fundamental laws. The irreversibility phenomenon may be explained only in 
transition to the formulation of the physical laws in terms of octonions. 
But in this most interesting case some of the theorems do not hold, on the 
basis of which the polarization density matrix is introduced and conclusions 
on the correspondence between tensors and bispinors are reached. The problem 
is to give an answer to the question: To what extent are those results for 
the correspondence between tensors and bispinors, which have been found for 
real matrix realizations of DM, valid for the octonion DM? 

This paper makes an attempt to give answers to the two above-formulated 
problems.

\section*{2. DM realization above a real field in Riemannian spaces of a dimension 
higher than four}

\bigskip

The Riemannian spaces of dimension $n \ge 4$ have been studied in connection 
with construction of matrix spaces (MS), that is the Riemannian spaces, in 
which the internal degrees of freedom properties are introduced through 
Dirac matrices $\gamma _{A} $. Subscripts $A,B,...$ take on values $A,B = 
1,2,...,n$, while the $\gamma _{A} $'s themselves are realized as square 
matrices $N \times N$ and satisfy relations

\begin{equation}
\label{eq3}
\gamma _{A} \gamma _{B} + \gamma _{B} \gamma _{A} = 2g_{AB} \cdot E
\end{equation}

Here$E$ is the unit matrix in the space of internal degrees of freedom.

The internal degrees of freedom are related, first, with transformations

\begin{equation}
\label{eq4}
\gamma _{A} \to {\gamma} '_{A} = S\left( {x} \right)\gamma _{A} S^{ - 
1}\left( {x} \right),
\end{equation}

\noindent
and, second, with the transition to Riemannian spaces of larger dimensions 
and different signatures. 

The MS theory in multidimensional Riemannian spaces with real realizations 
of DM is discussed in detail in refs. [1]-[3]. These papers also prove the 
following: 

The realization of DM above a real field frequently entails the notion of 
so-called maximum MS, in which the set of the quantities, the generatrices 
for which are DM, coincides with the set of all matrices of a given 
dimension. The maximum MS have odd dimension $n$.

\[
n = 2k + 1,
\]

\noindent
where $k$ is a positive integer. Their signature is therewith of form\\ 
$\left( {k + 1} \right)\left( { +}  \right)\& k\left( { -}  \right)$ or 
differs from that by a number of ``minuses'', which is a multiple of four.

The DM, which can be introduced in the Riemannian space possessing the above 
properties are square matrices $N \times N$, with $N$ relating to $k$ as

\[
N = 2^{k}.
\]

In any MS, either anti-Hermitizing matrix $D$ or Hermitizing matrix $C$ can 
be introduced. The $D$ or $C$ are determined as

\[
D\gamma _{a} D^{ - 1} = - \gamma _{a}^{ +}  ;\quad C\gamma _{a} C^{ - 1} = 
\gamma _{a}^{ +}  ;\quad a = 0,1,2,...,N - 1.
\]

The matrix $D$ or matrix $C$ can be used to introduce a Hermitean matrix 
set, whose existence, in its turn, is needed to introduce the concept of 
polarization density matrix.

\section*{3. Complex numbers and quaternions}

\bigskip

The results relating to determination of properties of those 
multidimensional Riemannian spaces, in which the real DM algebra is mapped 
isomorphously to the DM algebra in 4-dimensional space of signature $\left( 
{ - + + +}  \right)$ in realization of the latter above the real, complex 
and quaternionic fields are summarized in Table 1.

\bigskip

Table 1. Parameters characterizing the isomorphism between DM realized above 
different number fields and DM realized above the real number field

\bigskip

\newcommand{\PreserveBackslash}[1]{\let\temp=\\#1\let\\=\temp}
\let\PBS=\PreserveBackslash
\begin{longtable}
{|p{203pt}|p{114pt}|}
a & a  \kill
\hline
\multicolumn{2}{|p{317pt}|}{\textbf{A method for satisfaction 
of determining relation} $\gamma _{\alpha}  \gamma _{\beta}  + \gamma _{\beta}
\gamma _{\alpha}  = 2g_{\alpha \beta}  $\textbf{ with} 
$\alpha = 0,1,2,3$\textbf{ and signature} $
\left( { - + + +}  \right)$\textbf{}}  \\
\hline
\textbf{With the help of real matrices}& 
\textbf{With the help of matrices} $4 \times 4$\textbf{, but using different number bodies} \\
\hline
Matrices $4 \times 4$. \par 16 parameters.& 
Real number field \\
\hline
Subset of matrices $\left( {4 \times 4} \right) \otimes \left( {2 \times 2} \right)$. \par 32 parameters. \par In this realization method, an additional internal subspace of dimension 1 is actually introduced. \par Gauge group $U\left( {1} \right)$& 
Complex field. \par The transition to the matrix notation is performed using isomorphism \par \[
{\rm I} \Leftrightarrow E_{4 \times 4} \otimes i\sigma _{2} 
\]
 \\
\hline
Subset of matrices $\left( {4 \times 4} \right) \otimes \left( {4 \times 4} \right)$. \par 64 parameters. \par In this realization method, an additional internal subspace of dimension 3 is actually introduced. \par Gauge group $SU\left( {2} \right)$& 
Quaternion field. \par The transition to the matrix notation is performed using isomorphism \par \[
\left. {\begin{array}{l}
 {{\rm I}_{1} \Leftrightarrow E_{4 \times 4} \otimes i\rho _{2} \sigma _{1}}  \\ 
 {{\rm I}_{2} \Leftrightarrow E_{4 \times 4} \otimes i\sigma _{2}}  \\ 
 {{\rm I}_{3} \Leftrightarrow E_{4 \times 4} \otimes i\rho _{2} \sigma _{3}}  \\ 
 \end{array}}  \right\}
\]
 \\
\hline
\end{longtable}

As it follows from Table 1, the dimension of the internal space that appears 
in the transition from one number field to another is the same as the number 
of imaginary units in the number field. The Riemannian spaces are therewith 
subspaces of maximum MS.

\bigskip

\section*{4. Octonion DS}

\bigskip

Irrespective of the fact that octonions are discussed extensively in the 
literature (see, e.g., [4], [5]), nevertheless, here we present some 
information about these unusual numbers. Naturally, we will do this briefly 
and only to the extent, which is needed for the consistency of the 
discussion. 

The algebra of octonion imaginary units $\left\{ {e_{N}}  \right\}$ is 
determined as

\begin{equation}
\label{eq5}
e_{M} e_{N} = - \delta _{MN} e_{0} + C_{MNK} e_{K} 
\end{equation}

Here: $M,N,K = 1,2,...,7$; $C_{MNK} $ are quantities completely 
antisymmetric in their indices; nonzero components are:

\begin{equation}
\label{eq6}
C_{123} = C_{145} = C_{246} = C_{347} = C_{176} = C_{257} = C_{365} = 1.
\end{equation}

Quantity $\Delta \left[ {A,B,C} \right]$is called the associator of three 
octonions $A,B,C$:

\begin{equation}
\label{eq7}
\Delta \left[ {A,B,C} \right] = \frac{{1}}{{2}}\left\{ {\left( {AB} \right)C 
- A\left( {BC} \right)} \right\}.
\end{equation}

The whole specificity of the octonion algebra against the matrix algebra is 
that the associators (\ref{eq7}) are nonzero.

Perform the linear real transformation of symbols $e_{M} $ of the following 
form:

\begin{equation}
\label{eq8}
\left. {\begin{array}{l}
 {e_{M} \to {e}'_{M} = G_{MN} \cdot e_{N}}  \\ 
 {e_{0} \to {e}'_{0} = e_{0}}  \\ 
 \end{array}}  \right\}.
\end{equation}

Consider properties of tensor $G_{MN} $ in the 7-dimensional Euclidean 
space, in which the base vectors are symbols $e_{M} $. The substitution of 
${e}'_{M} $ into

\begin{equation}
\label{eq9}
{e}'_{M} {e}'_{N} = - \delta _{MN} e_{0} + C_{MNK} {e}'_{K} ,
\end{equation}

\noindent
which symbols ${e}'_{M} $ should satisfy, leads to the following two 
relations:

\begin{equation}
\label{eq10}
\left. {\begin{array}{l}
 {G_{MK} G_{NK} = \delta _{MN}}  \\ 
 {G_{MA} G_{NB} C_{ABC} = C_{MNS} G_{SC}}  \\ 
 \end{array}}  \right\}.
\end{equation}

Quantities $G_{MN} $ produce 14-parametric group $G_{2} $ of rank 2. 
According to the universally adopted classification, group $G_{2} $ is 
attributed to the exceptional Lie group category. Detailed information about 
the group $G_{2} $ can be found, e.g., in ref. [6]. 

It is known in advance that in the case of DS realization above the octonion 
body any isomorphous mapping of the appearing DS apparatus to the matrix 
apparatus cannot exist in principle. So the question is quite appropriate: 
Do the matrix realizations of DS have any bearing on the octonion DS 
whatsoever? 

To answer this question, make it our aim to construct the DS realization in 
the form of real DM in a multidimensional Riemannian space, which would 
satisfy the following requirement: 

$ \bullet $ When in algebraic operations with octonion DS $C_{ABC} $ play 
actually no role, the algebraic operations should map to the algebra of real 
DM of an appropriate dimension. This is true for the algebraic operations 
with DS $\left\{ {\gamma _{\alpha} }  \right\}$ near real DM $\left\{ {\bar 
{\gamma} _{\alpha} }  \right\}$. 

To meet this requirement, suppose that in the scheme under discussion there 
is the smallness parameter $0 < \lambda < < 1$, such that all matrix 
elements $\left( {\gamma _{\alpha}  - \bar {\gamma} _{\alpha} }  \right)$ 
modulo are of the order of $\lambda $. Write the matrices $\gamma _{\alpha}  
$ as

\begin{equation}
\label{eq11}
\gamma _{\alpha}  = \bar {\gamma} _{\alpha}  + f_{\alpha ;0} \cdot e_{0} + 
f_{\alpha ;N} \cdot e_{N} ;\quad \left( {N = 1,2,...,7} \right)
\end{equation}

Matrices (\ref{eq11}) will satisfy relation (\ref{eq1}), if small matrices $\left\{ 
{f_{\alpha ;0} ,f_{\alpha ;N}}  \right\}$ are of the form

\begin{equation}
\label{eq12}
f_{\alpha ;0} = \left[ {s_{0} ,\bar {\gamma} _{\alpha} }  \right]_{ -}  
;\quad f_{\alpha ;N} = \left[ {s_{N} ,\bar {\gamma} _{\alpha} }  \right]_{ - 
} ,
\end{equation}

\noindent
where $\left\{ {s_{0} ,s_{N}}  \right\}$are arbitrary small real matrices $4 
\times 4$. Upon substitution of (\ref{eq12}) into (\ref{eq11}) it turns out that octonion DS 
are written in the form

\begin{equation}
\label{eq13}
\gamma _{\alpha}  = \bar {\gamma} _{\alpha}  + \left[ {s_{0} ,\bar {\gamma 
}_{\alpha} }  \right]_{ -}  \cdot e_{0} + \left[ {s_{N} ,\bar {\gamma 
}_{\alpha} }  \right]_{ -}  \cdot e_{N} .
\end{equation}

The substitution of (\ref{eq13}) into (\ref{eq1}) shows that in the first smallness order 
the $C_{ABC} $ drop out and play no role. This means that the algebra with 
generatrices satisfying relation

\begin{equation}
\label{eq14}
\left[ {e_{M} ,\;e_{N}}  \right]_{ +}  = - 2\delta _{MN} e_{0} ,
\end{equation}

\begin{equation}
\label{eq15}
\left[ {e_{M} ,\;e_{N}}  \right]_{ -}  = 2C_{MNK} e_{K} ,
\end{equation}

\noindent
can be mapped in the first order of smallness to the algebra of real DM, in 
which instead of seven imaginary units $\left\{ {e_{N}}  \right\}$, seven 
matrix imaginary units $\left\{ {{\rm I}_{N}}  \right\}$ are used. The 
specific form of the real DM satisfying either above-formulated requirement 
can be as follows:

\begin{equation}
\label{eq16}
\left. {\begin{array}{l}
 {e_{1} \Leftrightarrow {\rm I}_{1} = E_{4 \times 4} \otimes i\rho _{2} 
\sigma _{1} \otimes \sigma _{1} \quad e_{2} \Leftrightarrow {\rm I}_{2} = 
E_{4 \times 4} \otimes i\sigma _{2} \otimes \sigma _{1} \quad}\\ 
e_{3}\Leftrightarrow {\rm I}_{3} = E_{4 \times 4} \otimes i\rho _{2} \sigma _{3} 
\otimes \sigma _{1} \quad \quad
 {e_{4} \Leftrightarrow {\rm I}_{4} = E_{4 \times 4} \otimes i\rho _{1} 
\sigma _{2} \otimes \sigma _{3} \quad}\\ 
e_{5} \Leftrightarrow {\rm I}_{5} = 
E_{4 \times 4} \otimes i\rho _{2} \otimes \sigma _{3} \quad e_{6} 
\Leftrightarrow {\rm I}_{6} = E_{4 \times 4} \otimes i\rho _{3} \sigma _{2} 
\otimes \sigma _{3}  \\ 
 {e_{7} \Leftrightarrow {\rm I}_{7} = E_{4 \times 4} \otimes E_{4 \times 4} 
\otimes i\sigma _{2} .} \\ 
 \end{array}}  \right\}
\end{equation}

The resultant multidimensional Riemannian space has dimension 11 and 
signature $1\left( { -}  \right)\& 10\left( { +}  \right)$. The DM in the 
space is written as:

\begin{equation}
\label{eq17}
\left. {\begin{array}{l}
 {\bar {\gamma} _{0} = - i\rho _{2} \sigma _{1} \otimes E_{4 \times 4} 
\otimes E_{2 \times 2} ;\quad}  
 {\bar {\gamma} _{1} = \rho _{1} \otimes E_{4 \times 4} \otimes E_{2 \times 
2}}\\
{\bar {\gamma} _{2} = \rho _{2} \sigma _{2} \otimes E_{4 \times 4} 
\otimes E_{2 \times 2} ;\quad \bar {\gamma} _{3} = \rho _{3} \otimes E_{4 
\times 4} \otimes E_{2 \times 2} ;} \\ 
 {\bar {\gamma} _{4} = i\rho _{2} \sigma _{3} \otimes i\rho _{2} \sigma _{1} 
\otimes \sigma _{1} ;\quad \bar {\gamma} _{5} = i\rho _{2} \sigma _{3} 
\otimes i\sigma _{2} \otimes \sigma _{1} }\\
{ \bar {\gamma} _{6} = i\rho 
_{2} \sigma _{3} \otimes i\rho _{2} \sigma _{3} \otimes \sigma _{1} \quad 
 {\bar {\gamma} _{7} = i\rho _{2} \sigma _{3} \otimes i\rho _{1} \sigma _{2} 
\otimes \sigma_{3}}}\\
{ \bar {\gamma} _{8} = i\rho _{2} \sigma _{3} 
\otimes i\rho _{2} \otimes \sigma _{3} ;\quad \bar {\gamma} _{9} = i\rho 
_{2} \sigma _{3} \otimes i\rho _{3} \sigma _{2} \otimes \sigma _{3} ;} \\ 
 {\bar {\gamma} _{10} = i\rho _{2} \sigma _{3} \otimes E_{4 \times 4} 
\otimes i\sigma _{2} .} \\ 
 \end{array}}  \right\}
\end{equation}

Expressions (\ref{eq13}) have the meaning of the ones in the first order of 
smallness for the first four DM among eleven DM. The expressions for all the 
eleven DM in the first order of smallness are derived from

\begin{equation}
\label{eq18}
\gamma _{\alpha}  = \bar {\gamma} _{\alpha}  + \left[ {s,\bar {\gamma 
}_{\alpha} }  \right]_{ -}  ;\quad \gamma _{N + 3} = \bar {\gamma} _{N + 3} 
+ \left[ {s,\bar {\gamma} _{N + 3}}  \right]_{ -}  ;\quad N = 1,2,...,7,
\end{equation}

\noindent
where

\begin{equation}
\label{eq19}
s = s_{0} \cdot E + s_{N} \cdot {\rm I}_{N} .
\end{equation}

Thus, in the linear approximation the octonion DM $\gamma _{\alpha}  $ can 
be treated as ordinary matrices, if for the basic matrices, in the vicinity 
of which the expansion proceeds, real DM are used in 11-dimensional 
Riemannian space of signature $1\left( { -}  \right)\& 10\left( { +}  
\right)$. Pay attention to the fact that except for the reality no other 
properties of DM in 11-dimensional Riemannian space have been used in this 
consideration. This means that instead of system (\ref{eq17}) that DM system can be 
used in the consideration, which has been derived in [7] from system (\ref{eq17}) 
through transition to the oblique-angled basis produced by simple root 
vectors of Lie algebra $E_{8} $.

In the general case the following rule remains valid: If it was possible to 
realize DS with the help of octonion DM $4 \times 4$, then after that one 
can transfer from one realization to another using transformations $G_{2} $.

\section*{5. Discussion}

Although the octonion DS can be written in the form of matrices $4 \times 4$ 
in the general case, but the algebra of the matrices possesses no 
associativity and, hence, cannot be mapped to the algebra of ordinary real 
matrices in a multidimensional space. In the linear approximation, however, 
the algebra of octonion DM is mapped to that of real DM in 11-dimensional 
Riemannian space of signature $1\left( { -}  \right)\& 10\left( { +}  
\right)$. One of possible DM systems in this space is of form (\ref{eq17}). 

The result obtained is of interest for several reasons.

\underline {Reason 1} is that the correspondence found by us between 
octonion DS and real DM in a multidimensional Riemannian space leads to the 
Riemannian space, in which the most stable vacuum state appears. Ref. [7] 
shows that the most stable vacuum state both among the internal subspaces of 
dimensions other than 7 and among DM of different spinor basis structure is 
the DM realization in the form of real matrices, in which the oblique-angled 
basis from the set of fundamental vectors of lattice $E_{8} $ is used. In 
this realization, the internal space dimension is 7; the specific form of 
the lattice DM is presented in ref. [7] and the matrix of transition from 
the orthonormal basis to the lattice one is given, e.g., in [7], [8]. 

\underline {Reason 2} is that using any number body, except for the octonion 
body, in physical theories does not allow us even to pose the question of 
explanation of the irreversibility of actual processes on the basis of 
time-reversible fundamental laws with writing the latter in terms of any 
number field. The irreversibility phenomenon may be explained only in 
transition to the formulation of the physical laws in terms of octonions. 

In this connection note that it has been long since the physicists have paid 
attention to the existence of an evident contradiction: on the one hand, the 
dynamic equations describing fundamental interactions possess time 
reversibility; on the other hand, actual processes that occur in the Nature 
are irreversible. R. Penrose in [9] writes: ``...It is hard to understand 
how our immense Universe could ``sink'' into one or another of the states 
with being unable to even imagine in what time direction to start! ...the 
only explanation ... remains: not all accurate physics laws are symmetric in 
time!...''. 

If DM are realized above the octonion body, then the transition amplitudes 
automatically cease to be associative. While this just means that the 
reversibility in time does disappear at the level of fundamental processes 
in the microworld. In fact, if $A_{1} ,A_{2} ,A_{3} ,...$ are amplitudes of 
the transitions from initial state $t_{0} $ to states arising at times 
$t_{1} ,t_{2} ,t_{3,...} $, then the amplitude for one of the paths of the 
process proceeding in the time-forward direction should be found according 
to rule

\begin{equation}
\label{eq20}
\left( {\left( {A_{1} \cdot A_{2}}  \right) \cdot A_{3}}  \right)...,
\end{equation}

\noindent
while the conjugate amplitude for the process running in the time-backward 
direction should be found according to rule

\begin{equation}
\label{eq21}
\left( {A_{1} \cdot \left( {A_{2} \cdot A_{3} ...} \right)} \right).
\end{equation}

At the level of real, complex and quaternionic numbers expressions (\ref{eq20}), 
(\ref{eq21}) lead to the same probabilities of transitions. But as soon as octonions 
come into use, the equality between expressions (\ref{eq20}), (\ref{eq21}), generally 
speaking, disappears. Moreover, the body of octonion numbers is the only one 
possessing this property. This means that we may necessarily resort to the 
octonion quantities for explanation of the irreversibility of processes.

The above considerations and results justify the multiple attempts to 
consider the octonion wave functions for half-integer spin particles. We 
only point out to refs. [5], [10], [11] as typical papers from the 
standpoint of the method for consideration of octonion Dirac matrices. The 
method of these papers is valid only to the quadratic approximation, as in 
these papers there is either explicit or implicit transition to so-called 
split octonions (introduction of the outer imaginary unit commutating with 
all octonions) or the octonion composition rule is replaced by the open 
product. Similar (or equivalent) techniques restore the associativity of the 
modified number body and allow the standard matrix apparatus to be employed. 
However, in so doing a most interesting part of the octonion specificity is 
lost.

A method for description of the half-integer spin particle dynamics is the 
method of mapping of tensors to bispinors developed in a number of papers 
(see, e.g., [3]). In the method, one of principal objects is bispinor matrix 
$Z$. For the octonion implementation of DS, the matrix $Z$ exists in the 
linear approximation and, as it follows from (\ref{eq13}), coincides with $S^{ - 
1}$. Through multiplication on the right by the projectors, states with 
different quantum numbers can be separated from the bispinor matrix. For 
example, one of the subgroups of group $G_{2} $ is $SU\left( {3} \right)$. 
In the general case there is no bispinor matrix, however, the results 
obtained using the methods for consideration of the transformations of DM $4 
\times 4$, which are suggested in ref. [12], remain valid. 

Thus, the vacuum stability requirement can be made consistent with using the 
most general number body. In so doing any violation of the bounds of the 
7-dimensional internal Euclidean space will result in vacuum instability 
(and appearance of tachyons as a consequence).

The work was carried out under partial financial support by the 
International Science and Technology Center (ISTC Project \#1655).

\bigskip

\section*{References}

\bigskip

[1] M.V. Gorbatenko, A.V. Pushkin. \textit{VANT; Ser.: Teor. i Prikl. Fiz}. 
\textbf{1(\ref{eq1})}, 49 (1984).

[2] M.V. Gorbatenko. \textit{TMF}. \textbf{103}, \textbf{1}, 32 (1995).

[3] M.V. Gorbatenko, A.V. Pushkin. \textit{VANT; Ser.: Teor. i Prikl. Fiz}. 
\textbf{2-3}, 61 (2000).

[4] B.A. Rozenfeld. \textit{Non-Euclidean geometries.} Moscow. Gostekhizdat 
Publishers (1955).

[5] D.F. Kurdgelaidze. \textit{Introduction to nonassociative classical 
field theory}. Tbilisi. Metsnnegeva (1987).

[6] R.E. Behrends, J. Dreitlein, C. Fronsdal, W. Lee. \textit{Rev. Mod. 
Phys.,} \textbf{34}, No. 1, 1 (January 1962). 

[7] M.V. Gorbatenko, A.V. Pushkin. \textit{Physical Vacuum Properties and 
Internal Space Dimension.} gr-qc/0409095. To be published in GRG.

[8] J.H. Conway, N.J.A. Sloane. \textit{Sphere Packings, Lattices and 
Groups.} Springer-Verlag. New York (1988).

[9] R. Penrose. \textit{Singularities and asymmetry in time.} In: ``The 
General Relativity''. Moscow, Publishers (1983), p. 233.

[10] Sirley Marques-Bonham. \textit{The Dirac equation in a non-Riemannian 
manifold III: An analysis using the algebra of quaternions and octonions.} 
J. Math. Phys. \textbf{32} (\ref{eq5}), 1383 (1991).

[11] S. Margues, C.G. Oliveira.\textit{ An extension of quaternionic metrics 
to octonions.} J. Math. Phys. \textbf{26 (\ref{eq12})}, 3131 (1985).

[12] N.D. Sen Gupta. \textit{On the Invariance Properties of the Dirac 
Equation.} Nuovo Cimento, Vol. XXXVI, N.4 (1965).

\end{document}